\documentclass{article}
\usepackage{ijcai25}

\usepackage{times}
\usepackage{soul}
\usepackage{url}
\usepackage[hidelinks]{hyperref}
\usepackage[utf8]{inputenc}
\usepackage[small]{caption}
\usepackage{graphicx}
\usepackage{amsmath}
\usepackage{amsthm}
\usepackage{booktabs}
\usepackage{algorithm}
\usepackage{amsmath}
\usepackage{algorithm}
\usepackage{algorithmic}
\usepackage{amssymb}
\usepackage{algorithmic}
\usepackage[switch]{lineno}

\title{Generating Large Semi-Synthetic Graphs of Any Size}

\author{
Rodrigo Tuna$^1$
\and
Carlos Soares$^{1,2,3}$\\
\affiliations
$^1$Faculdade de Engenharia, Universidade do Porto\\
$^2$Artificial Intelligence \& Computer Science Lab.(LIACC -- member of LASI LA), Universidade do Porto\\
$^3$Fraunhofer AICOS\\
\emails
\{up201904967, csoares\}@fe.up.pt
}

\begin{document}

\maketitle

\begin{abstract}
Graph generation is an important area in network science. Traditional approaches focus on replicating specific properties of real-world graphs, such as small diameters or power-law degree distributions. Recent advancements in deep learning, particularly with Graph Neural Networks, have enabled data-driven methods to learn and generate graphs without relying on predefined structural properties. Despite these advances, current models are limited by their reliance on node IDs, which restricts their ability to generate graphs larger than the input graph and ignores node attributes. To address these challenges, we propose Latent Graph Sampling Generation (LGSG), a novel framework that leverages diffusion models and node embeddings to generate graphs of varying sizes without retraining. The framework eliminates the dependency on node IDs and captures the distribution of node embeddings and subgraph structures, enabling scalable and flexible graph generation. Experimental results show that LGSG performs on par with baseline models for standard metrics while outperforming them in overlooked ones, such as the tendency of nodes to form clusters. Additionally, it maintains consistent structural characteristics across graphs of different sizes, demonstrating robustness and scalability.
\end{abstract}

\section{Introduction}
 Graph Generation has been a well-studied topic in the field of Network Science~\cite{gen:barabasi-network}. Classic generative algorithms focused on generating synthetic graphs that exhibited some properties of real-world graphs that had been previously observed, such as small diameter or power-law degree distribution~\cite{gen:r-mat,gen:ws}. The advances in deep learning, specifically in Graph Neural Networks~\cite{gnn-book}, extended deep learning to non-euclidean data. This lead to new graph generation approaches that resort to data-driven algorithms that do not focus on graph properties known \emph{a priori} but rather learn to capture the semantic meaning of a graph or a dataset of graphs to generate new ones. These methods have been applied to code generation~\cite{gen:code-gen}, molecule generation~\cite{gen:MolGAN}, and to extract semantic graphs from text~\cite{gen:semantic-graph}.

However, current deep learning methods face limitations in generating graphs larger than the input graph. This constraint arises because models that focus on generating one large graph rely on node IDs~\cite{gen:GraphRNN,gen:NetGAN,gen:SaGess} to identify and generate nodes within a given input graph. Consequently, it is not feasible to produce graphs with more nodes than the input graph, as each node must be assigned a unique ID, and duplicate IDs are not permissible. Notably, these node IDs do not encode any meaningful information about the graph's structure or the relationships between nodes; they merely impose a fixed node ordering during the generation process. Additionally, these approaches often disregard potential node attributes, focusing exclusively on the graph's structural properties, which further limits their expressive capacity.

We introduce a novel framework, Latent Graph Sampling Generation (LGSG), that leverages the strengths of diffusion models; which are well-established in image generation~\cite{diff:beat-gans} and increasingly effective in generating graph-structured data~\cite{gen:DiGress}. Unlike traditional approaches, our framework enables the generation of graphs with more nodes than the input graph, overcoming the limitations of fixed-size output graphs. Moreover, the presented framework is able to generate graphs with different sizes without the need for retraining, offering unprecedented flexibility in adapting to diverse size constraints in graph generation. By learning node embeddings that encode structural and relational information, the model captures the distribution of embeddings and subgraph structures, enabling scalable and cohesive graph generation. 

We evaluate our approach against classic graph generative models by comparing graph statistics between input and generated graphs, demonstrating its ability to produce diverse and structurally meaningful graphs while supporting size variability. Results indicate that the presented framework is on par with the baseline models when comparing metrics that these algorithms are designed to capture. The framework also outperforms the baselines in metrics the other algorithms do not encompass in the generation, e.g. Cluster Coefficient. In addition, we provide a study on the scalability of the algorithm by using the same model, trained on an input graph, to generate graphs of different sizes. Concluding that, although the graph metrics of graphs reproduced by some baselines change less with respect to size, our framework maintains consistency and generates graphs with realistic structural characteristics regardless of scale.

This paper starts by documenting graph generation. We then present a recent diffusion graph generative model, SaGess~\cite{gen:SaGess}, that focuses on generating large graphs. Then, in Section~\ref{sec:alg}, we characterize the proposed framework. Finally, the experimental setup is explained in Section~\ref{sec:es}, and the results are discussed in Section~\ref{sec:res}. The code used for this work is available at \url{https://github.com/rodrigotuna/LGSG}

\section{Background}\label{sec:back}
In this section, we then present the main concepts related to our work. We start by briefly detailing the problem of graph generation and then present the graph diffusion model SaGess~\cite{gen:SaGess}, on which we base our architecture.
\subsection{Graph Generation}
A graph $\mathcal{G}(\mathcal{V}, \mathcal{E})$, with a set of $N = |\mathcal{V}|$ nodes $\mathcal{V}$ and a set of edges $\mathcal{E}$, where $(u,v) \in \mathcal{E}$ indicates an edge between nodes $u, v \in \mathcal{V}$.  In the context of machine learning, as graph datasets include features from each node, a graph is represented as $G(X,E)$ with adjacency matrix $E \in \mathbb{R}^{N \times N}$, where $E_{i,j} = 1$ if there is an edge between nodes $v_i$ and $v_j$ and is $0$ otherwise; and node features $X \in \mathbb{R}^{N \times d}$ where $x_i$ is a vector of $d$ features  $\forall 0 \le i \le N$. The problem of graph generation is formally defined as, given a dataset of graphs $D\{\mathcal{G}_i\}$, possibly one, originated from an underlying data distribution $p$, we want to generate another dataset of graphs $D\{\mathcal{G}'_i\}$ such that are sampled from the distribution $p$~\cite{gen:form}.

Graph generation is the process of creating new graph-structured data that simulates the structure and properties of real-world graphs. The topic was first studied in the mathematical field of Random Graphs~\cite{gen:rand-graph}, intersecting probability theory and graph theory. It is also a fundamental topic of Network Science~\cite{gen:barabasi-network,gnn-book}. This task is essential for applications that require artificial yet realistic network representations, such as social networks~\cite{gen:ba}, molecular structures~\cite{gen:MPGVAE}, knowledge graphs extraction from text~\cite{gen:semantic-graph}, or code generation~\cite{gen:code-gen}. The existing solutions for this problem sample from probabilistic models that capture the statistical properties of existing graphs, allowing the generation of synthetic or semi-synthetic graphs with similar characteristics.

Classical methods for graph generation focus on creating probabilistic models that capture fundamental properties observed in real-world networks, such as degree distribution, clustering coefficients, and average path lengths. Notable examples include the Erdős–Rényi model~\cite{gen:er}, which assumes edges are formed independently with a fixed probability, and the Barabási–Albert model~\cite{gen:ba}, which generates scale-free networks using a preferential attachment mechanism. These approaches often fall short of capturing more complex patterns, such as hierarchical structures or overlapping communities, commonly observed in real-world graphs and are not used to reproduce a specific example. Contrarily, the R-MAT model~\cite{gen:r-mat} and Kronecker model~\cite{gen:kronecker} allow for fitting a small number of parameters to an input graph using algorithms like KronFit.

With advances in deep learning, particularly through graph neural networks (GNNs), more expressive models have emerged, enabling the generation of graphs that better reflect the complexities of real-world data. 
The use of variational autoencoders~\cite{gen:graphvae,gen:MPGVAE}, recurrent neural networks~\cite{gen:GraphRNN}, generative adversarial networks (GANs)~\cite{gen:MMGAN,gen:NetGAN,gen:SHADOWCAST} and diffusion-based models~\cite{gen:DiGress,gen:SaGess} have been adapted to graph-structured data.  
Deep learning-based graph generation operates in two main settings: learning from a single graph and learning from a dataset of graphs. In the single-graph setting~\cite{gen:SaGess,gen:NetGAN}, models are trained on a large, complex graph, such as a social network, to extract structural patterns like motifs, communities, or hierarchies, enabling the generation of graphs that closely mirror the original. In contrast, dataset-based learning~\cite{gen:DiGress,gen:GraphRNN} involves generalizing across multiple smaller graphs, such as molecular structures, by capturing graph-level properties and the overall distribution of the dataset.
These state-of-the-art approaches address challenges such as capturing multi-scale structures and learning from specific real-world graph datasets. They allow for generated graphs to capture properties that are not explicitly encoded into the algorithm design. However, these methods suffer from scalability issues that prevent them from generating large-scale graphs. 

Evaluating the effectiveness of these generative models remains a key challenge~\cite{gen:met-obray}, with several approaches developed to assess how well these models produce realistic and meaningful graphs. Metrics commonly used for evaluation include structural similarity measures, which quantify how closely the generated graphs align with real-world graph statistics. These measures are instantiated as comparisons of degrees, clustering coefficients, and orbit counts~\cite{gen:GraphRNN} between the generated and real graphs. The evaluation can also take into account task-specific criteria, common in the domains of chemistry or biology. In the specific context of molecule generation, the quality is measured using metrics that assess the chemical properties of the molecules, such as synthetic accessibility score and drug-likeness score. 

\subsection{SaGess}
SaGess~\cite{gen:SaGess} is a diffusion-based framework designed for scalable graph generation, where a single large-scale input graph $\mathcal{G}$ serves as the basis for generating new semi-synthetic graph instances. SaGess is composed of three main steps: (1) A graph dataset $D(\mathcal{G}) = \{G_1, ..., G_m\}$ is produced by sampling subgraphs from the initial graph. Each subgraph $G^i$, is composed by an adjacency matrix, and a matrix of one-hot enconding of the node IDs in the original graph.  (2) These subgraphs are then used to train the DiGress model, which generates small synthetic graph samples $D'(\mathcal{G}) = \{G'_1, ..., G'_M\}$ each generated graph includes a matrix of one-hot encoding of the node IDs in the synthetic graph.  (3) Finally, the model systematically merges these small graphs into a single large graph by performing the graph union of the subgraphs, $\hat{\mathcal{G}} = \bigcup_{i=0}^M \mathcal{G}'_i$, using the generated node IDs as node identification in the synthetic graph. This agglomeration process is repeated until the number of edges in the synthetic graph equals the number of edges of the original graph. This approach allows it to generate large, real-world networks more efficiently. Results indicate that SaGess outperformed other deep graph generative methods in a number of selected graph descriptive measures and tasks like link prediction.

\section{Algorithm}\label{sec:alg}
\begin{figure}
    \centering
    \includegraphics[width=\columnwidth]{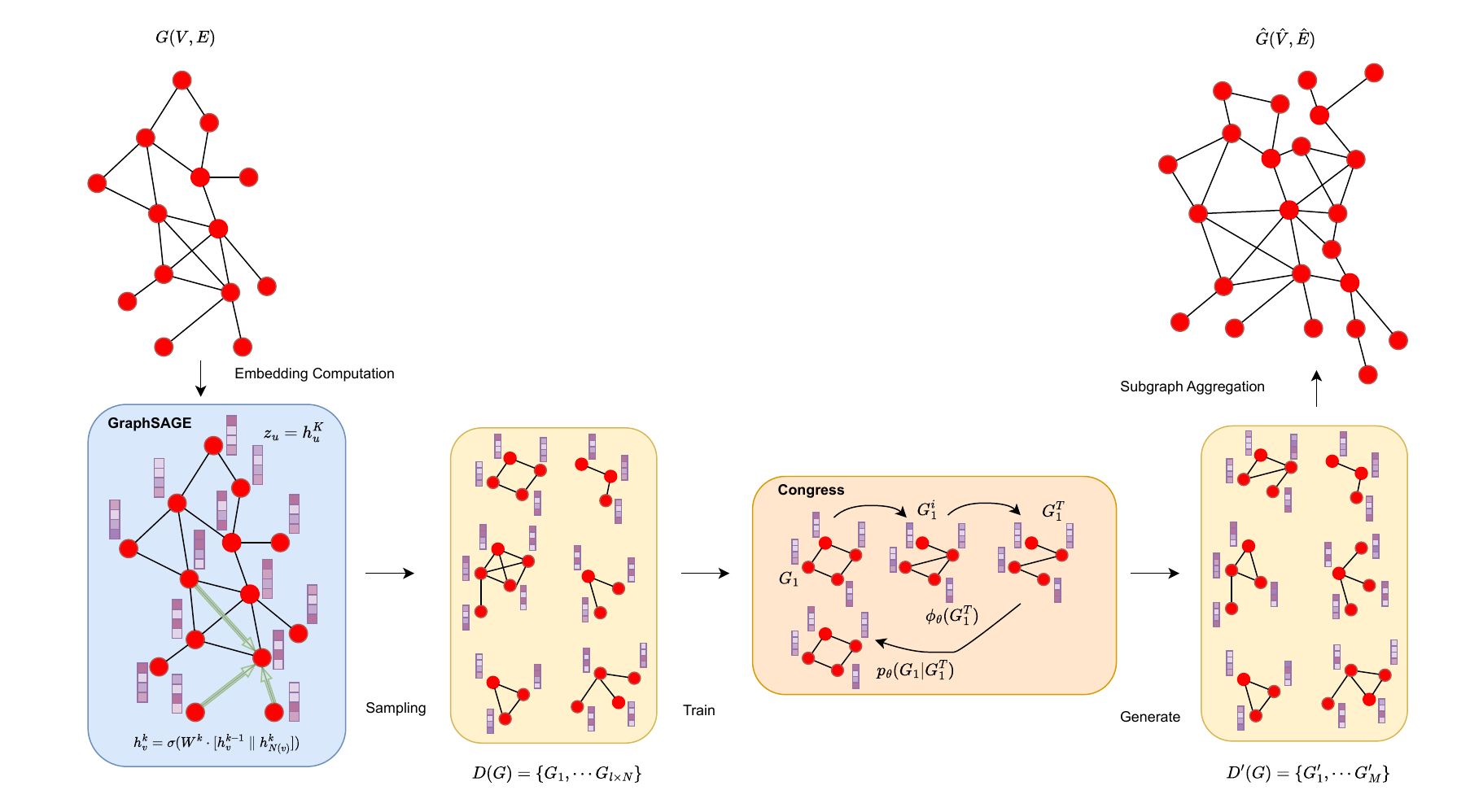}
    \caption{Diagram detailing the LGSG framework.}
    \label{fig:LS}
\end{figure}

We propose a new learning framework, Latent Graph Sampling Generation (LGSG), that is able to generate graphs with more nodes than the number of nodes of the input graph, making use of node embeddings. 
The model has four steps: first, a node embedding is calculated for each of the nodes in the graph; then a dataset of subgraphs is created by sampling from the graph via random walks; a diffusion model is trained on the dataset; synthetic subgraphs are sampled from the model and aggregated into an output graph using two algorithms that are described below (Schema presented in Figure~\ref{fig:LS}). The model is then able to learn the distribution of node embeddings as well as the subgraph structure. Each synthetic subgraph is comprised of the subgraph structure along with the generated node embeddings. These embeddings are used by the aggregation algorithms that are built on the assumption that similar nodes have a small embedding distance. The use of node embeddings instead of node IDs allows the model to learn more meaningful representations of the data. Moreover, methods that utilize node IDs constrain the output graph to have, at most, the same number of nodes as the input graph. The graphs generated by our method can have any number of nodes, which is parameterizable. 

\subsection{Learning Graph Representation}
Node embeddings are vectors associated with each node that convey information regarding the graph structure and potential node attributes. The rationale behind embeddings is to distill the high dimensionality into a vector of reduced dimension. Node embeddings not only reflect the structural properties of the graph but also incorporate node-specific information, thereby providing a rich and informative representation of each node within the graph. To compute these embeddings, we utilize the GraphSAGE~\cite{GraphSage} model under a self-supervised learning strategy. The algorithm operates under the assumption that nodes tend to connect to similar nodes, so the model computes similar embeddings between connected nodes while at the same time producing dissimilar embeddings between distant nodes.

Given the input graph $\mathcal{G}(\mathcal{V},\mathcal{E})$, together with a matrix $X$ of node attributes, the GraphSAGE framework begins by initializing the embedding for each node $v \in \mathcal{V}$ with its input features, such that $h_v^0 = x_v$. For each layer $k \in \{1, \cdots, K\}$, the model samples a fixed-size set of neighbors $Neigh(v)$ for each node $v$. The embeddings of these sampled neighbors are then aggregated using a function denoted as $\text{AGGREGATE}^k$. For the experiments of this paper, we use the mean as the $\text{AGGREGATE}$; see the original paper~\cite{GraphSage} for a list of other possible functions. This aggregation can be represented mathematically as $h_{Neigh(v)}^k = \text{AGGREGATE}^k(\{h_u^{k-1} | u \in Neigh(v)\})$. The node’s embedding from the previous layer is then concatenated with the aggregated embedding and passed through a non-linear transformation, formulated as $h_v^k = \sigma(W^k \cdot [h_v^{k-1} \parallel h_{Neigh(v)}^k])$, $W^k$ is a learnable weight matrix, and $\sigma$ is a non-linear activation function.

The final embeddings are obtained from the output of the last layer $K$, represented as $z_v = h_v^K$, for node $v \in \mathcal{V}$. The GraphSAGE algorithm is trained by minimizing a loss function that ensures nearby nodes have similar embeddings, sampled by fixed-sized random-walks; and the embeddings of distant nodes to be distinct, negatively sampled at random. The loss function for a node $u \in V$ is given by:
\begin{equation}
    l(\mathcal{G}, u) = -\log(\sigma(z_u \cdot z_v)) - Q \cdot \mathbb{E}_{v_n \sim P_n(v)} [\log(\sigma(-z_u \cdot z_{v_n}))]
\end{equation}
where $v$ whose distance from $u$ is smaller than a fixed length, $Q$ is the number of negative samples, $P_n$ is a negative sampling distribution, and $\sigma$ is the sigmoid function.

The embeddings from the final layer of the GraphSAGE model are then extracted as the output.

\subsection{Sampling a Subgraph Dataset}
Since only one large graph observation is available, we first need to divide the graph into smaller substructures that can be used to train a model.
After the computation of the matrix of node embeddings $Z \in  \mathrm{R}^{N \times d}$ where $d$ is the embedding size. A random walk of length $m$ is defined as $W = \{v_{w_1}, v_{w_2}, \dots, v_{w_m} \}$, where the next node is chosen based on the probability vector $P_i = e_i / \text{deg}(v_i)$ when currently at node $v_i$. Here, $e_i$ is the $i$-th row of the adjacency matrix $E$. For each node $v_i \in V$, we initiate a specified number of $l$ random walks, ensuring that every node appears in the dataset at least $l$ times. Consequently, the subgraph dataset is defined as $D(\mathcal{G}) = \{ \mathcal{G}_0, \dots, \mathcal{G}_{l \times N} \}$, where each subgraph $G_i$ is derived from the walk $W_i$ and its edges $(W_i \times W_i) \cap \mathcal{E}$. For our machine learning application, each subgraph $G_i$ is represented by its adjacency matrix $E_i \in \mathbb{R}^{m \times m}$, and node embeddings $Z_i \in \mathbb{R}^{m \times d}$, where $\{{z_i}_j\}$ is the row of the embeddings matrix, $Z$ corresponding to node $v_{w_j}$ in the original graph. The use of embeddings significantly reduces the memory complexity of the algorithm. The size of the feature matrices is smaller when compared to the version proposed in the previous chapter. The memory complexity is $\mathcal{O}(m \times d)$ instead of $\mathcal{O}(N\times m)$ for the hot-one encoding of size $N$. This makes the algorithm more scalable as it does not depend on the size of the input graph. 

\subsection{Subgraph Generation}
We wish to learn a model that learns the subgraph distribution so that we can sample synthetic subgraphs from it along with the respective node embeddings.
We use a continuous model for modelling the subgraph distribution, given that the embeddings are a continuous variable. We use ConGress~\cite{gen:DiGress} to learn from the generated dataset of subgraphs. Given a graph $G = (Z, E)$ The algorithm adds Gaussian noise independently to each node and edge such that:

\begin{equation}
     q(Z^{t} \mid Z^{t-1}) = \mathcal{N}(\alpha^{t|t-1} Z^{t-1}, (\sigma^{t|t-1})^2 I) 
\end{equation}
\begin{equation}
     q(E^{t} \mid E^{t-1}) = \mathcal{N}(\alpha^{t|t-1} E^{t-1}, (\sigma^{t|t-1})^2 I)
\end{equation}
This is equivalent to:
\begin{equation}
    q(Z^{t} \mid Z^{t-1}) = \mathcal{N}(\alpha^{t|t-1} Z^{t-1}, (\sigma^{t|t-1})^2 I)
\end{equation}
\begin{equation}
   q(E^{t} \mid E^{t-1}) = \mathcal{N}(\alpha^{t|t-1} E^{t-1}, (\sigma^{t|t-1})^2 I)
\end{equation}

where $\alpha^{t|t-1} = \frac{\alpha^{t}}{\alpha^{t-1}}$ and $(\sigma^{t|t-1})^2 = (\sigma^{t})^2 - (\alpha^{t|t-1})^2 (\sigma^{t-1})^2$.

A neural network $\phi(\theta,G^t)$ is trained to predict the noise added to the samples $\hat{\epsilon}_X$, which relates to the noise parameters as follows:
\begin{equation}
       \alpha^{t} \hat{Z} = X^{t} - \sigma^{t} \hat{\epsilon}_X
   \text{ and, }
   \alpha^{t} \hat{E} = E^{t} - \sigma^{t} \hat{\epsilon}_E
\end{equation}
The loss function is the mean squared error between the added and predicted noises. 

The denoising process can be computed by:
\begin{equation}
   q(Z^{t-1} \mid Z, Z^{t}) = \mathcal{N}(\mu^{t \to t-1}(Z, Z^{t}), (\sigma^{t \to t-1})^2 I)
\end{equation}

where $\mu^{t \to t-1}(Z, Z^{t}) = \alpha^{t|t-1} \left(\frac{(\sigma^{t-1})^2}{\sigma^2_t}\right) Z^{t} + \alpha^{t-1} \left(\frac{(\sigma^{t|t-1})^2}{(\sigma^{t})^2}\right) Z$, and $\sigma^{t \to t-1} = \frac{\sigma^{t|t-1} \sigma^{t-1}}{\sigma^{t}}$, similar formulas apply for the edges, $E$.

A discrete sample for the generated graph is computed by taking the argmax $E^0$.

\subsection{Aggregating the Final Graph}
\begin{figure}
    \centering
    \includegraphics[width=\columnwidth]{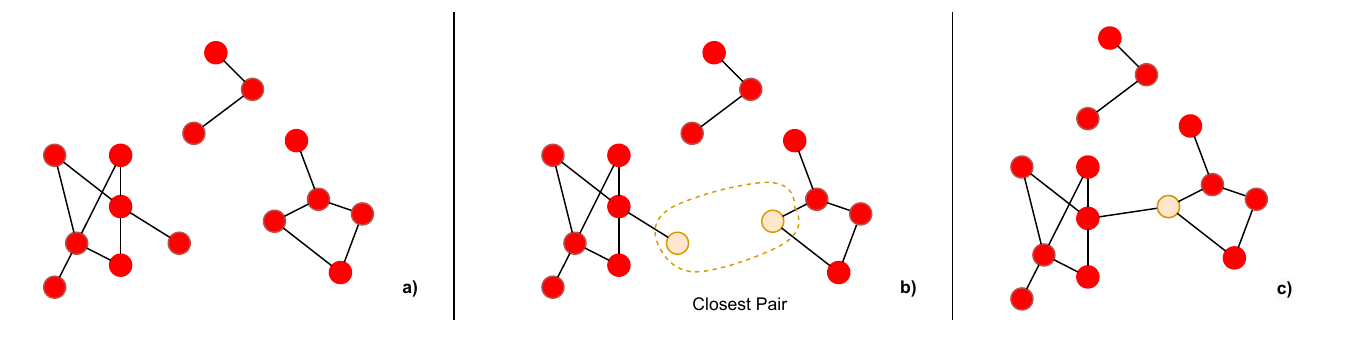}
    \caption{Toy example of one iteration of the Node Aggregation Algorithm for subgraph assembling. a) The current generated graph. b) The closest pair of nodes in terms of embedding distance is chosen and nodes are joined in a supernode c) The resulting generated graph after the iteration.}
    \label{fig:NA}
\end{figure}

\begin{figure}
    \centering
    \includegraphics[width=\columnwidth]{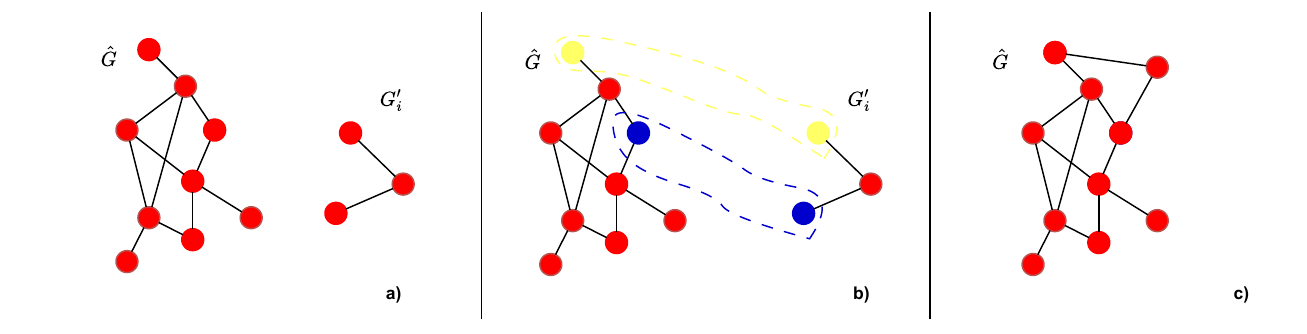}
    \caption{Toy example of one iteration of the Threshold Matching Algorithm for subgraph assembling. a) The current generated graph and the new sample. b) The matching between nodes in the sampled subgraph and the generated graph. c) The resulting generated graph}
    \label{fig:TS}
\end{figure}

We propose two different approaches for assembling the output graph given the generated subgraph samples. Both approaches are parameterized to control the output size. The Node Aggregation algorithm constructs the final graph by iteratively merging the nodes from different sampled subgraphs with the smallest embedding distance. The algorithm is parametrized by the number of nodes. The Threshold Matching algorithm is parameterized by a threshold that indicates the maximum distance such that nodes are considered sufficiently similar and iteratively adds the nodes of the sampled subgraphs given the threshold.

The Node Aggregation Algorithm ~\ref{alg:node_agg} starts with the union of all sampled subgraphs $\hat{\mathcal{G}} =\bigcup \mathcal{G}'_i$, which creates a disjoint graph of $M$ connected components. At each iteration, the nodes with the smallest embedding distance are merged until we reach the desired number of nodes, as portrayed in an example in Figure~\ref{fig:NA}. 
Formally, given a dataset of generated graphs $D'(\mathcal{G}) = \{\mathcal{G}'_1(\mathcal{V}'_1,\mathcal{E}'_1), \cdots, \mathcal{G}'_M(\mathcal{V}'_M,\mathcal{E}'_M)\}$ and a desired number of nodes $\nu$, we define the set of generated nodes $\mathcal{V}' = \bigcup \mathcal{V}'_i$. We assign each node as its own super node $sup(v_i) = v_i \forall v_i \in V'$. At each iteration, the two nodes, which are not part of the same subgraph with the smallest distance between embeddings $argmin \{|u - v| | \forall u,v \in V, \nexists i: u,v \in \mathcal{V}'_i\}$, are used by joining their super nodes $sup(v_i) = sup(u), \forall v_i \in \mathcal{V}': sup(v_i) = sup(v)$. This process is repeated until we arrive at the desired number of nodes $|\{sup(v_i): \forall v_i \in \mathcal{V}'\}| = \nu$, and we form the graph $\hat{\mathcal{G}}=(\hat{\mathcal{V}} = \{sup(v_i)\ | \forall v_i \in \mathcal{V}'\}, \hat{\mathcal{E}} = \{sup(u),sup(v) | \forall (u,v) \in \bigcup \mathcal{E}'_i\})$. The algorithm makes use of a disjoint union set to track the super nodes and a priority queue to retrieve the closest pair of embeddings and has a $\mathcal{O}((M\times d)^2)$ running time due to the calculation of distances between all pairs of nodes.

\begin{algorithm}
\caption{Node Aggregation}\label{alg:node_agg}
\begin{algorithmic}[1]
\STATE \textbf{Input}: $D'(\mathcal{G}) = \{\mathcal{G}'_1(\mathcal{V}'_1,\mathcal{E}'_1), \cdots, \mathcal{G}'_M(\mathcal{V}'_M,\mathcal{E}'_M)\}$, $\nu$
\STATE $pairs = \{(u,v), key= |u-v| | \forall u,v \in\mathcal{V}'| \nexists i : u,v \in \mathcal{V}'_i\}$
\STATE $heapify(pairs)$
\STATE $sup(\mathcal{V}') = dsu()$
\STATE $nodes = |\mathcal{V}'|$
\WHILE{$nodes \ne \nu$}
    \STATE $u,v = pairs.pop()$
    \IF{$sup.same(u,v)$}
        \STATE continue
    \ELSE 
    \STATE $sup.link(u,v)$
    \STATE $nodes-=1$
    \ENDIF
\ENDWHILE
\STATE $\hat{\mathcal{G}}=(\hat{\mathcal{V}}=\{sup(v_i) | \forall v_i \in \mathcal{V}'\}, \hat{E} = \{sup(u),sup(v) | \forall (u,v) \in \bigcup \mathcal{E}'_i\})$
\STATE \textbf{Output}: $\hat{\mathcal{G}}$
\end{algorithmic}
\end{algorithm}

The Threshold Matching Algorithm~\ref{alg:threshold},  constructs the graph by iteratively adding each sample to the output graph. We start with an empty graph with no nodes or edges and iteratively construct it by adding new subgraph samples. For each subgraph sample, we match every node in the subgraph to the closest existing node in the current generated graph if the distance between the embeddings of the two nodes falls under a given threshold. Otherwise, we add a new node to the current generated graph. The edges of the subgraph are added to the output graph as edges between the matched nodes. Figure~\ref{fig:TS}
Formally, given a dataset of generated graphs $D'(\mathcal{G}) = \{\mathcal{G}'_1(\mathcal{V}'_1,\mathcal{E}'_1), \cdots, \mathcal{G}'_M(\mathcal{V}'_M,\mathcal{E}'_M)\}$ and a threshold $\delta$, we start with an empty graph $\hat{\mathcal{G}}(\hat{\mathcal{V}} = \emptyset, \hat{\mathcal{E}} = \emptyset)$. For each subgraph $\mathcal{G}'_i$ we assign to every node $v_j \in \mathcal{V}'_i$ from the current generated graph or the node itself, such that:
\begin{equation}
    node(v_j) = 
\begin{cases} 
argmin(|u - v_j|) \forall u \in \hat{\mathcal{V}} & \text{if }  |u-v_j| \le \delta\\
v_j & \text{otherwise}
\end{cases} 
\end{equation}
We then add the new nodes and edges to $\hat{G}$ so we get $\hat{\mathcal{G}}(\hat{\mathcal{V}} = \hat{\mathcal{V}} \cup \{node(v_j)\}, \forall v_j \in \mathcal{V}'_i,  \hat{\mathcal{E}} = \hat{\mathcal{E}} \cup \{(node(u), node(v)\}, \forall (u,v) \in \mathcal{E}'_i)$. The algorithm ends when all the subgraphs of the dataset $D'(\mathcal{G})$ have been added, and the output is $\hat{\mathcal{G}}$. The implementation of the algorithm is straightforward, and the running time is again $\mathcal{O}((M\times d)^2)$.

\begin{algorithm}
\caption{Threshold Matching}\label{alg:threshold}
\begin{algorithmic}[1]
\STATE \textbf{Input}: $D'(\mathcal{G}) = \{\mathcal{G}'_1(\mathcal{V}'_1,\mathcal{E}'_1), \cdots, \mathcal{G}'_M(\mathcal{V}'_M,\mathcal{E}'_M)\}$, $\delta$
\STATE $\hat{\mathcal{G}}(\hat{\mathcal{V}} = \emptyset, \hat{\mathcal{E}} = \emptyset)$
\FOR{$i \in {1, \cdots, M}$}
    \STATE $node = \emptyset$
    \FOR{$v_j \in \mathcal{V}'_i$}
    \STATE $u = argmin \{|u_i-v_j| | \forall u_i \in \hat{\mathcal{V}}\}$
    \IF{$|u-v_j| \le \delta$}
        \STATE $nodes(v_j) = u$
    \ELSE
        \STATE $nodes(v_j) = v_j$
    \ENDIF
    \ENDFOR
    \STATE $\hat{\mathcal{V}}=\hat{\mathcal{V}}\cup\{node(v_j)\},\forall v_j \in \mathcal{V}'_i$
    \STATE $\hat{\mathcal{E}} = \hat{\mathcal{E}} \cup \{(node(u), node(v)\}, \forall (u,v) \in \mathcal{E}'_i$
\ENDFOR
\STATE \textbf{Output}: $\hat{\mathcal{G}}(\hat{\mathcal{V}},\hat{\mathcal{E}})$
\end{algorithmic}
\end{algorithm}
\section{Experimental Setup} \label{sec:es}
We empirically evaluate our algorithm and compare it to other graph generative methods using real-world datasets. Since our work is the first deep-learning model that is able to generate graphs larger than the input graph, we resort to classical models for comparison. We compare the algorithms in terms of graph quality by computing several graph statistics at different levels of size increase. We perform 5 runs of each algorithm for each size and dataset. 

\subsection{Baselines}
We select 3 different relevant baselines with different types of generation strategies. 

\paragraph{Erdos-Renyi}~\cite{gen:er}: A random graph model where each pair of nodes is connected with a fixed probability independently of other pairs. We fit the edge probability as the edge density of the original graph.
\paragraph{Barabasi-Albert}~\cite{gen:ba}: A scale-free network model where new nodes preferentially attach to existing nodes with higher degrees, leading to a power-law degree distribution. We fit the attachment parameter as half of the average degree.
\paragraph{Kronecker}~\cite{gen:kronecker}: A recursive graph model where a small initiator matrix is repeatedly Kronecker-multiplied to generate a large graph, often resulting in self-similar and hierarchical structures. We use the KronFit algorithm to fit the initiator matrix to the original graph.
\paragraph{GenCAT}~\cite{GenCAT}: A graph generator that samples from learned latent factors. It can capture relationships between node labels, attributes, and graph topology, including phenomena like core/border structures and homophily/heterophily.

\subsection{Datasets}
    We evaluate the algorithm on 4 different real-world fairness datasets, detailed in Table~\ref{tab:graph_description4} from the torch geometric package:
\paragraph{CiteSeer}~\cite{CiteSeer}: A citation network of indexed academic literature.
\paragraph{Cora}~\cite{Cora}: A citation network of machine learning papers.
\paragraph{Facebook}~\cite{Facebook}: A friend network of Facebook users.
\paragraph{Wiki}~\cite{Wiki}: A data set from Wikipedia pages and hyperlinks between them.

    \begin{table}[t]
        \caption{Description of the datasets}
        \centering
        \begin{tabular}{cccc}
        Dataset & Nodes & Edges \\ 
        \hline
            CiteSeer & 3,327 & 4,552 \\
            Cora & 2,708 & 5,278  \\
            Facebook & 1,045 & 27,755 \\
            Wiki  & 2,277 & 31,421 \\
        \end{tabular}
        \label{tab:graph_description4}
    \end{table}
    
\subsection{Metrics}
We measure the ability of the model to capture the properties of the input graph and reproduce them when scaling the input graph. We compute different statistics in the original graph and compare them to the same measures on synthetic graphs at different scales. The properties of interest are the metrics used in~\cite{gen:SaGess,gm:fair-gan} that are independent of graph size:
\paragraph{Average Degree}: Average value of node degrees. Helps understand overall connectivity.
\paragraph{Edge Distribution Entropy (EDE)}: The relative entropy of the degree distribution. Highlights uniformity or presence of hubs.
\paragraph{Gini}: The Gini coefficient of the degree distribution. Indicates inequality in node connectivity.
\paragraph{Clustering Coefficient}: The average value of node cluster coefficients. Reflects local cohesiveness or community structure.
\paragraph{Assortativity}: The correlation between the degrees of connected nodes. Reveals mixing patterns of similar or dissimilar nodes.

\section{Results}\label{sec:res}
We present the results obtained by using the two algorithms to produce graphs using the same subgraph samples and compare each algorithm individually with the baseline. This section addresses the following research questions (RQs):
\paragraph{RQ1}: Can the proposed algorithm produce graphs with metrics similar to the input graphs?
\paragraph{RQ2}: How do the metrics vary across different sizes of generated graphs?
\paragraph{RQ3}: Which graph aggregation method is superior for graph generation?

\subsection{Node Aggregation}

\begin{figure}
    \includegraphics[width=\columnwidth]{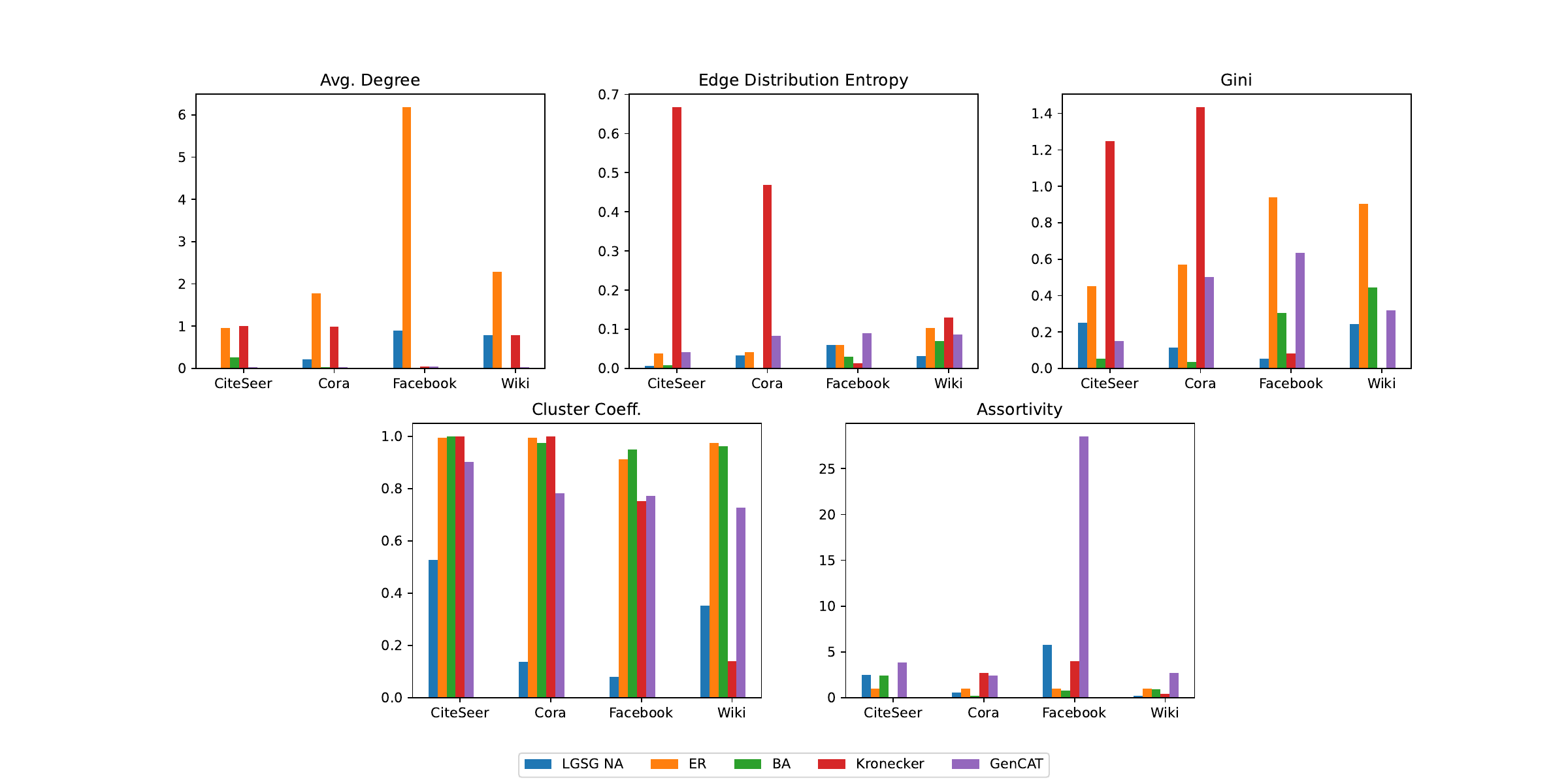}
    \caption{Comparison of the LGSG model using the Node Aggregation algorithm and baselines using performance metrics.}
    \label{fig:latentna}
\end{figure}

Figure~\ref{fig:latentna} shows the relative distance between the average of the measures of the graphs across different sizes and the same measures computed for the original graph. The Node Aggregation algorithm is shown to perform better than the baselines. It can be seen that, especially in the clustering coefficient, where methods have a similar performance to that of a random graph, our method is able to correctly capture the cohesiveness and the community structure of the graph. The Barabasi-Albert and the GenCAT methods are designed to explicitly control the average node degree through the added edges parameter and are superior to our method. However, they are inferior to our method in recreating the degree distribution as exhibited by the EDE and Gini Coefficient metrics. The Barabasi-Albert model struggles especially in the Wiki dataset, which does not represent a social network. Our method fails to capture the degree distribution in the Facebook dataset given it exhibited worse performance in the average node degree and the EDE. The reason for this behavior is that the dataset represents an ego graph, and one of the nodes skews the average degree of the graph; thus, our model is not able to recreate that node. However, the graphs generated by our model can still have similar Gini indexes. All methods struggle with the assortativity of the graph on all datasets except Wiki.

\begin{figure}
    \includegraphics[width=\columnwidth]{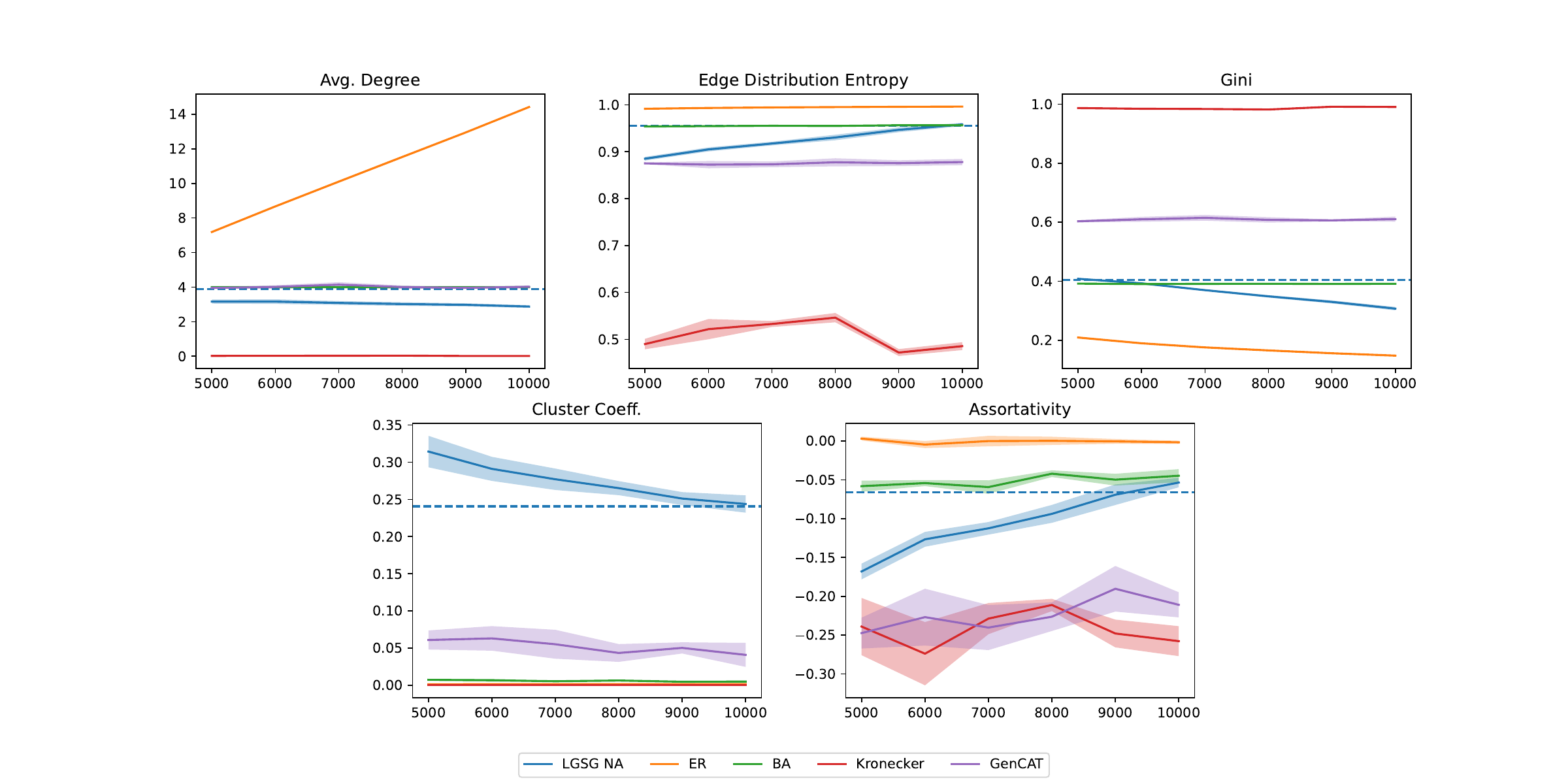}
    \caption{Evolution of graph measures with the size of generated graphs for the Cora dataset.}
    \label{fig:latentnacora}
\end{figure}

Assessing the impact that the node parameter has on the quality of the generated graphs, we show in Figure~\ref{fig:latentnacora} the measure of the different metrics for generated graphs of different sizes for the Cora dataset. We can observe that metrics for our model fluctuate more than other methods, meaning that the number of nodes impacts the quality of the generated graph. It suggests that the number of generated subgraphs from the model should be chosen accordingly. We also identify that although a great stochastic component exists in our algorithm, the variation in most properties is low.  

\subsection{Threshold Matching}
\begin{figure}
    \includegraphics[width=\columnwidth]{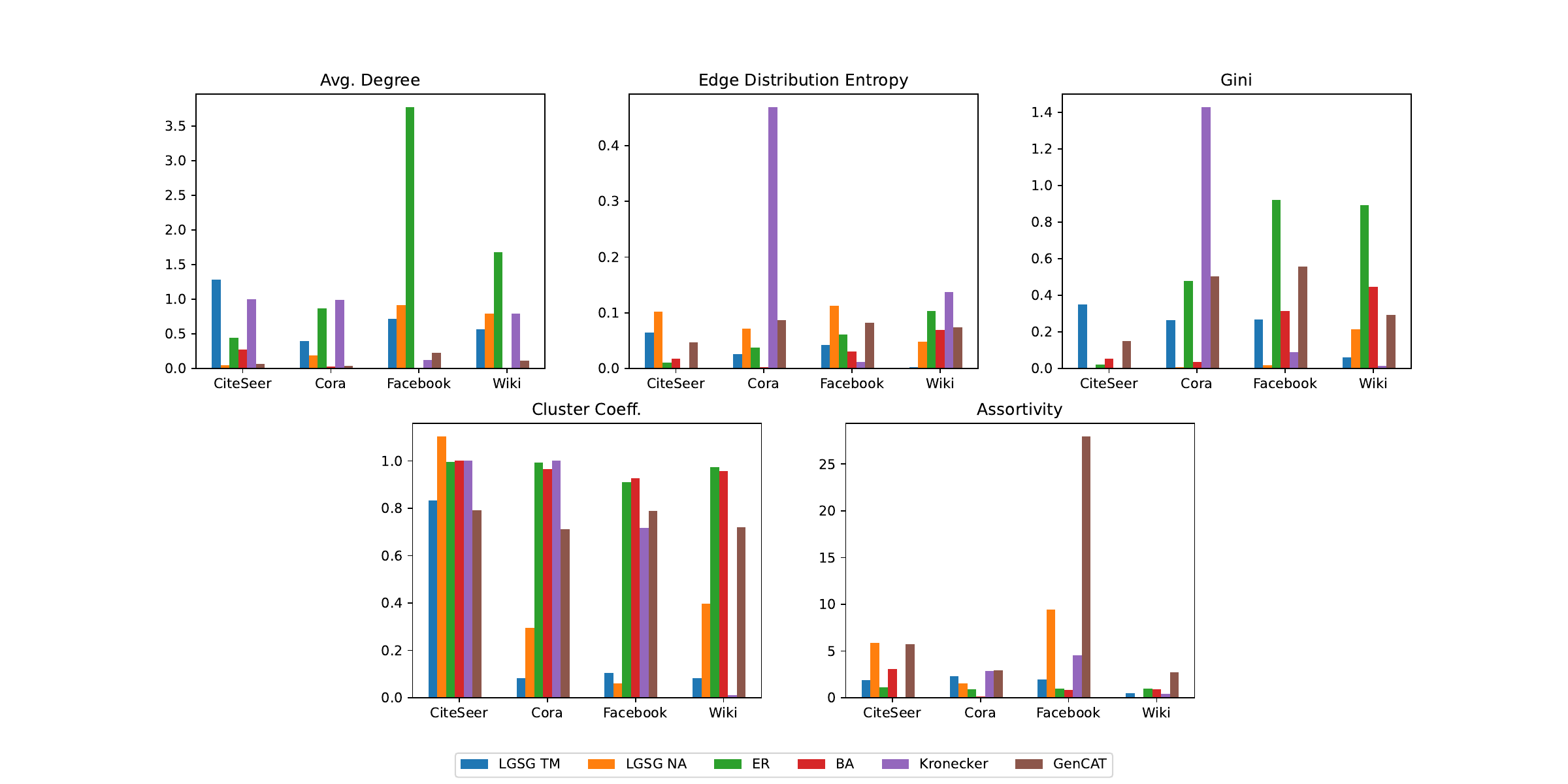}
    \caption{Comparison of the LGSG model using the Threshold Matching algorithm and baselines using performance metrics.}
    \label{fig:latent}
\end{figure}

Figure~\ref{fig:latent} shows the relative distance between the average of the measures of the graphs across different sizes and the same measures computed for the original graph. It can be seen that this method overall improves the performance over the baseline. Although this method struggles to recreate the average node degree and the Gini coefficient of the degree distribution, it can still learn certain aspects of the degree distribution given the values for the edge distribution entropy. This method is superior in exhibiting assortativity when compared to the Node Aggregation algorithm and performs better in datasets where the other method fails. The clustering coefficient of graphs generated by the method is closer to the clustering coefficient of the input graph when compared with graphs generated by the baselines.

\begin{figure}
    \includegraphics[width=\columnwidth]{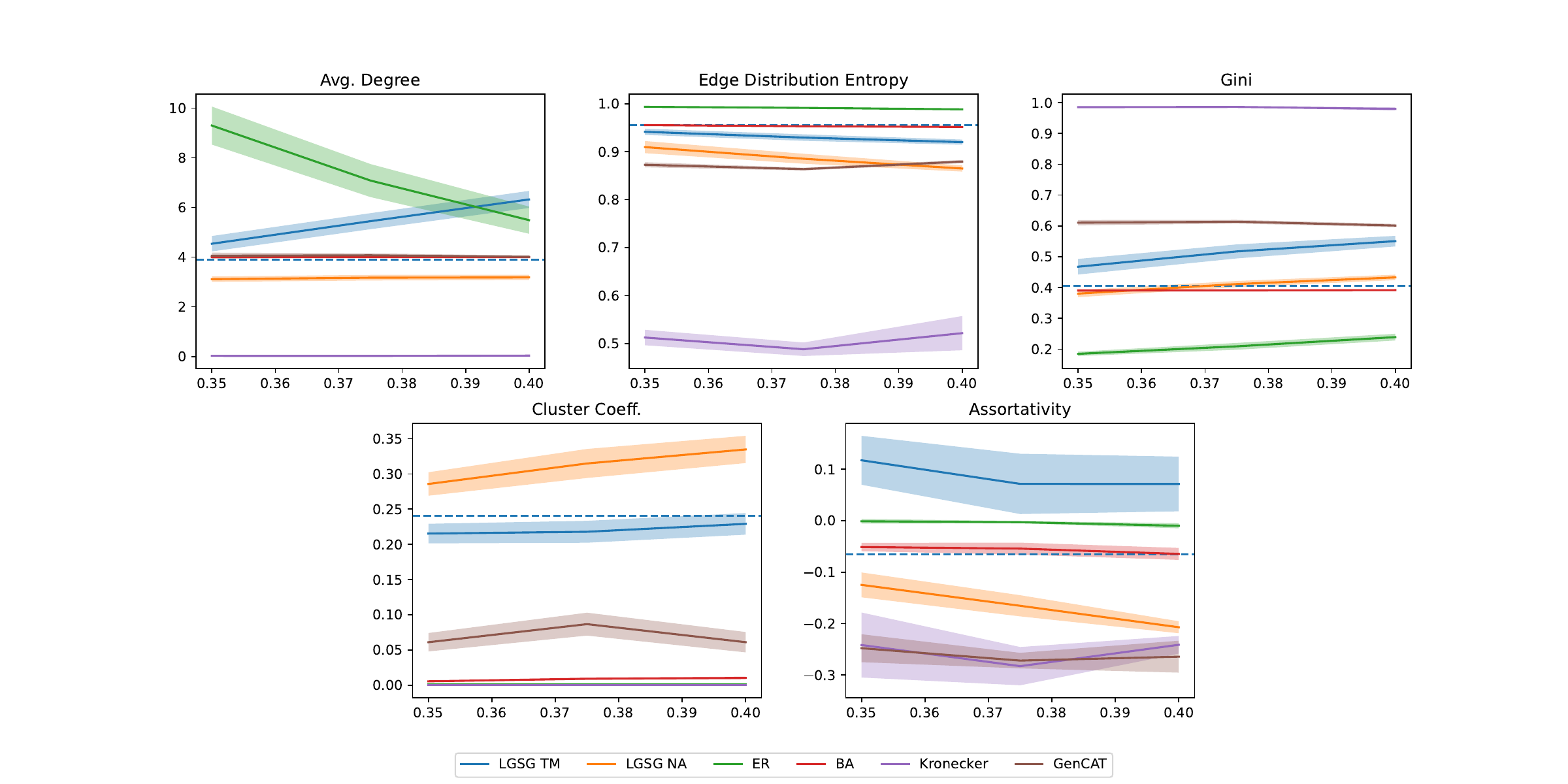}
    \caption{Evolution of graph measures with the threshold parameter of generated graphs for the Cora dataset.}
    \label{fig:latentnacoraths}
\end{figure}

Figure~\ref{fig:latentnacoraths} shows the different graph measures for graphs generated by the Threshold Matching algorithm with varying threshold parameters on the Cora dataset. The graphs are compared with graphs generated by the other methods with the same number of nodes. It is observable that the graph metrics vary more compared to other methods as the number of nodes increases. This indicates that the chosen threshold affects the quality of the generated graph. The threshold is a sensitive parameter and not perfectly indicative of the size of the generated graph. Small changes to the threshold and even the same threshold lead to graphs of very distinct sizes. This is exhibited by the high variation of metrics in both our models and the baseline models, which did not occur when we were defining the number of nodes. Moreover, the threshold is not consistent across different datasets, as the same threshold produces graphs of completely different relative sizes. In the CiteSeer dataset, with the same threshold, the output graphs were smaller than the original graph, which was not the case in the other three datasets.

\subsection{Discussion}
\begin{figure}
    \includegraphics[width=\columnwidth]{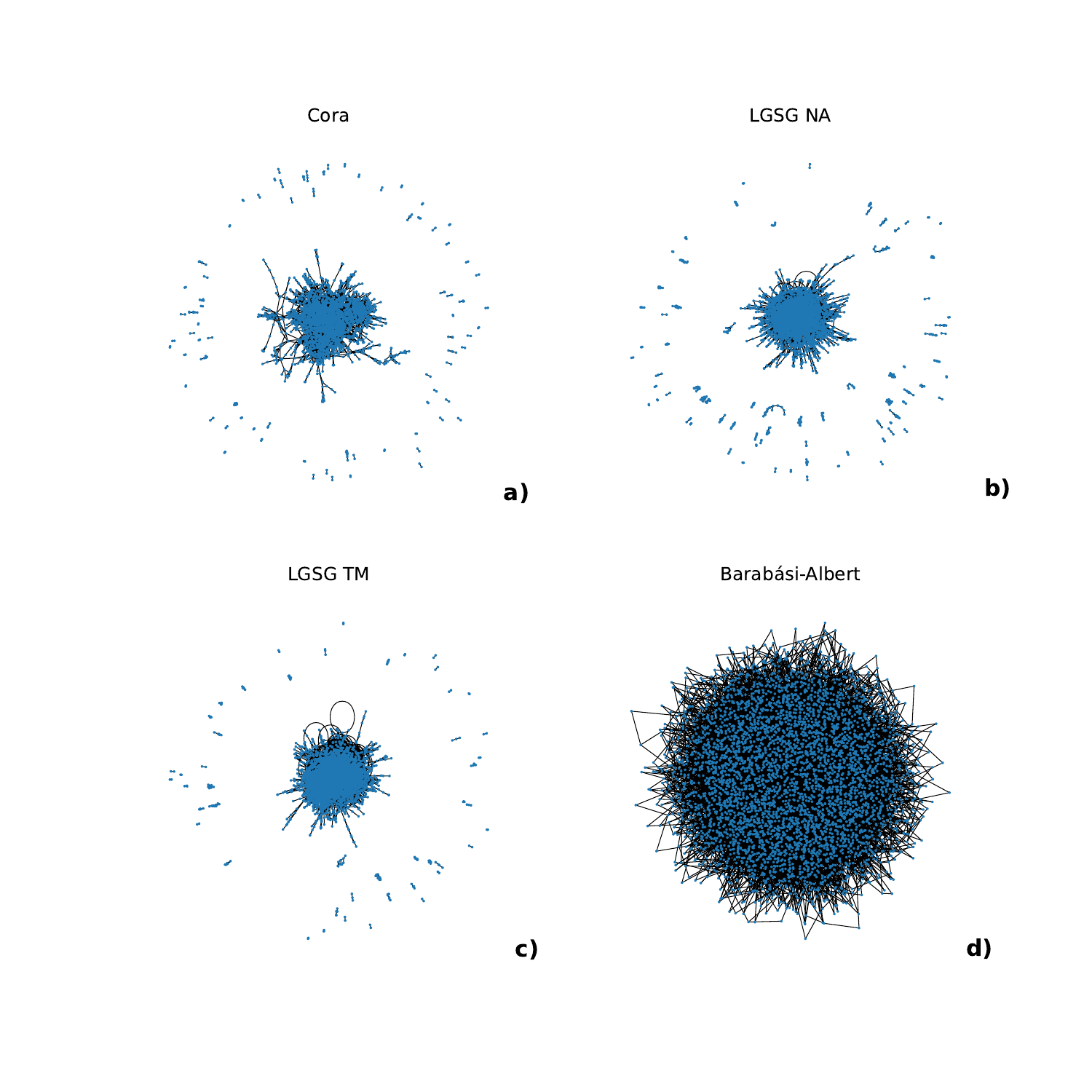}
    \caption{Visual comparison between the Cora graph a) and larger generated graphs by the LGSG framework b),c) and the Barabasi-Albert algorithm d)}
    \label{fig:viz_comp}
\end{figure}

When comparing the two proposed methods for output graph assembling, we conclude there is no significant difference between the two methods in terms of the overall quality of generation, as shown in Figure~\ref{fig:latent}. However, a distinction between the two methods is that choosing the number of outputs makes the generation more controllable. The threshold is a sensible parameter, and it is not clear what an appropriate threshold should be, as it is dependent on the input dataset. Figure~\ref{fig:viz_comp} shows a visual comparison of the original Cora graph and the generated graphs using our method at an increase of around double the number of nodes. We can see that both variants of the LGSG model capture the structural properties, namely the various connected components exhibited in the Cora graph. Conversely, other models can not recreate such fine-grained properties of graphs.

\section{Conclusion}
We proposed LGSG, a novel framework for graph generation that overcomes key limitations of existing methods, including the reliance on fixed-size input graphs and node IDs. By leveraging diffusion models and learning node embeddings, our framework enables the generation of graphs with varying sizes without retraining while preserving structural and relational properties. 

The results demonstrate that the proposed approach performs comparably to baseline models in capturing graph metrics they implicitly address while outperforming them in metrics the baselines are not designed to handle, such as clustering coefficients. This result highlights that deep learning models do not need to be specifically designed to capture specific graph properties. Furthermore, the scalability study highlights the robustness of the framework in maintaining consistent structural characteristics across graphs of different sizes. The visual analysis of generated graphs confirms that our method effectively captures fine-grained properties, such as connected components, which traditional models do not reproduce, hinting at a superiority when using synthetic data to train a machine learning model or to simulate a process on a synthetic network.

However, this work does not address certain challenges that remain open for future exploration, such as conditioning the generation to produce graphs with desired properties and extending the algorithm to generate node attributes alongside the graph structure.

\bibliographystyle{named}
\bibliography{ijcai25}

\begin{thebibliography}{}

\bibitem[\protect\citeauthoryear{Barabási and Albert}{1999}]{gen:ba}
Albert-László Barabási and Réka Albert.
\newblock Emergence of scaling in random networks.
\newblock {\em Science}, 286(5439):509–512, October 1999.

\bibitem[\protect\citeauthoryear{Barabási and Pósfai}{2016}]{gen:barabasi-network}
Albert-László Barabási and Márton Pósfai.
\newblock {\em Network science}.
\newblock Cambridge University Press, Cambridge, 2016.

\bibitem[\protect\citeauthoryear{Bojchevski \bgroup \em et al.\egroup }{2018}]{gen:NetGAN}
Aleksandar Bojchevski, Oleksandr Shchur, Daniel Zügner, and Stephan Günnemann.
\newblock Netgan: Generating graphs via random walks, 2018.

\bibitem[\protect\citeauthoryear{Bollobás}{2001}]{gen:rand-graph}
Béla Bollobás.
\newblock {\em Random graphs}.
\newblock Number~73 in Cambridge studies in advanced mathematics. Cambridge University Press, 2 edition, 2001.

\bibitem[\protect\citeauthoryear{Brockschmidt \bgroup \em et al.\egroup }{2019}]{gen:code-gen}
Marc Brockschmidt, Miltiadis Allamanis, Alexander~L. Gaunt, and Oleksandr Polozov.
\newblock Generative code modeling with graphs, 2019.

\bibitem[\protect\citeauthoryear{Cao and Kipf}{2022}]{gen:MolGAN}
Nicola~De Cao and Thomas Kipf.
\newblock Molgan: An implicit generative model for small molecular graphs, 2022.

\bibitem[\protect\citeauthoryear{Chakrabarti \bgroup \em et al.\egroup }{2004}]{gen:r-mat}
Deepayan Chakrabarti, Yiping Zhan, and Christos Faloutsos.
\newblock {R-MAT:} {A} recursive model for graph mining.
\newblock In Michael~W. Berry, Umeshwar Dayal, Chandrika Kamath, and David~B. Skillicorn, editors, {\em Proceedings of the Fourth {SIAM} International Conference on Data Mining, Lake Buena Vista, Florida, USA, April 22-24, 2004}, pages 442--446. {SIAM}, 2004.

\bibitem[\protect\citeauthoryear{Chen \bgroup \em et al.\egroup }{2018}]{gen:semantic-graph}
Bo~Chen, Le~Sun, and Xianpei Han.
\newblock Sequence-to-action: End-to-end semantic graph generation for semantic parsing, 2018.

\bibitem[\protect\citeauthoryear{Cohen-Karlik \bgroup \em et al.\egroup }{2024}]{gen:form}
Edo Cohen-Karlik, Eyal Rozenberg, and Daniel Freedman.
\newblock Overcoming order in autoregressive graph generation, 2024.

\bibitem[\protect\citeauthoryear{Dhariwal and Nichol}{2021}]{diff:beat-gans}
Prafulla Dhariwal and Alex Nichol.
\newblock Diffusion models beat gans on image synthesis, 2021.

\bibitem[\protect\citeauthoryear{Erdos and Renyi}{1960}]{gen:er}
Paul Erdos and Alfred Renyi.
\newblock On the evolution of random graphs.
\newblock {\em Publ. Math. Inst. Hungary. Acad. Sci.}, 5:17--61, 1960.

\bibitem[\protect\citeauthoryear{Flam-Shepherd \bgroup \em et al.\egroup }{2021}]{gen:MPGVAE}
Daniel Flam-Shepherd, Tony~C Wu, and Alan Aspuru-Guzik.
\newblock Mpgvae: improved generation of small organic molecules using message passing neural nets.
\newblock {\em Machine Learning: Science and Technology}, 2(4):045010, jul 2021.

\bibitem[\protect\citeauthoryear{Gamage \bgroup \em et al.\egroup }{2019}]{gen:MMGAN}
Anuththari Gamage, Eli Chien, Jianhao Peng, and Olgica Milenkovic.
\newblock Multi-motifgan (mmgan): Motif-targeted graph generation and prediction, 2019.

\bibitem[\protect\citeauthoryear{Giles \bgroup \em et al.\egroup }{1998}]{CiteSeer}
C.~Lee Giles, Kurt~D. Bollacker, and Steve Lawrence.
\newblock Citeseer: an automatic citation indexing system.
\newblock In {\em Proceedings of the Third ACM Conference on Digital Libraries}, DL '98, page 89–98, New York, NY, USA, 1998. Association for Computing Machinery.

\bibitem[\protect\citeauthoryear{Hamilton \bgroup \em et al.\egroup }{2018}]{GraphSage}
William~L. Hamilton, Rex Ying, and Jure Leskovec.
\newblock Inductive representation learning on large graphs, 2018.

\bibitem[\protect\citeauthoryear{Leskovec and Mcauley}{2012}]{Facebook}
Jure Leskovec and Julian Mcauley.
\newblock Learning to discover social circles in ego networks.
\newblock In F.~Pereira, C.J. Burges, L.~Bottou, and K.Q. Weinberger, editors, {\em Advances in Neural Information Processing Systems}, volume~25. Curran Associates, Inc., 2012.

\bibitem[\protect\citeauthoryear{Leskovec \bgroup \em et al.\egroup }{2009}]{gen:kronecker}
Jure Leskovec, Deepayan Chakrabarti, Jon Kleinberg, Christos Faloutsos, and Zoubin Ghahramani.
\newblock Kronecker graphs: An approach to modeling networks, 2009.

\bibitem[\protect\citeauthoryear{Limnios \bgroup \em et al.\egroup }{2023}]{gen:SaGess}
Stratis Limnios, Praveen Selvaraj, Mihai Cucuringu, Carsten Maple, Gesine Reinert, and Andrew Elliott.
\newblock Sagess: Sampling graph denoising diffusion model for scalable graph generation, 2023.

\bibitem[\protect\citeauthoryear{Maekawa \bgroup \em et al.\egroup }{2023}]{GenCAT}
Seiji Maekawa, Yuya Sasaki, George Fletcher, and Makoto Onizuka.
\newblock Gencat: Generating attributed graphs with controlled relationships between classes, attributes, and topology, 2023.

\bibitem[\protect\citeauthoryear{McCallum \bgroup \em et al.\egroup }{2000}]{Cora}
Andrew~Kachites McCallum, Kamal Nigam, Jason Rennie, and Kristie Seymore.
\newblock Automating the construction of internet portals with machine learning.
\newblock {\em Inf. Retr.}, 3(2):127–163, July 2000.

\bibitem[\protect\citeauthoryear{O'Bray \bgroup \em et al.\egroup }{2022}]{gen:met-obray}
Leslie O'Bray, Max Horn, Bastian Rieck, and Karsten Borgwardt.
\newblock Evaluation metrics for graph generative models: Problems, pitfalls, and practical solutions, 2022.

\bibitem[\protect\citeauthoryear{Rozemberczki \bgroup \em et al.\egroup }{2019}]{Wiki}
Benedek Rozemberczki, Carl Allen, and Rik Sarkar.
\newblock Multi-scale attributed node embedding, 2019.

\bibitem[\protect\citeauthoryear{Simonovsky and Komodakis}{2018}]{gen:graphvae}
Martin Simonovsky and Nikos Komodakis.
\newblock Graphvae: Towards generation of small graphs using variational autoencoders, 2018.

\bibitem[\protect\citeauthoryear{Tann \bgroup \em et al.\egroup }{2021}]{gen:SHADOWCAST}
Wesley Joon-Wie Tann, Ee-Chien Chang, and Bryan Hooi.
\newblock Shadowcast: Controllable graph generation, 2021.

\bibitem[\protect\citeauthoryear{Vignac \bgroup \em et al.\egroup }{2023}]{gen:DiGress}
Clement Vignac, Igor Krawczuk, Antoine Siraudin, Bohan Wang, Volkan Cevher, and Pascal Frossard.
\newblock Digress: Discrete denoising diffusion for graph generation, 2023.

\bibitem[\protect\citeauthoryear{Wang \bgroup \em et al.\egroup }{2023}]{gm:fair-gan}
Zichong Wang, Charles Wallace, Albert Bifet, Xin Yao, and Wenbin Zhang.
\newblock $\mathrm fg^2an$: Fairness-aware graph generative adversarial networks.
\newblock In Danai Koutra, Claudia Plant, Manuel Gomez~Rodriguez, Elena Baralis, and Francesco Bonchi, editors, {\em Machine Learning and Knowledge Discovery in Databases: Research Track}, pages 259--275, Cham, 2023. Springer Nature Switzerland.

\bibitem[\protect\citeauthoryear{Watts and Strogatz}{1998}]{gen:ws}
Duncan~J. Watts and Steven~H. Strogatz.
\newblock Collective dynamics of ‘small-world’ networks.
\newblock {\em Nature}, 393(6684):440--442, 1998.

\bibitem[\protect\citeauthoryear{Wu \bgroup \em et al.\egroup }{2022}]{gnn-book}
Lingfei Wu, Peng Cui, Jian Pei, and Liang Zhao.
\newblock {\em Graph Neural Networks: Foundations, Frontiers, and Applications}.
\newblock Springer Singapore, Singapore, 2022.

\bibitem[\protect\citeauthoryear{You \bgroup \em et al.\egroup }{2018}]{gen:GraphRNN}
Jiaxuan You, Rex Ying, Xiang Ren, William~L. Hamilton, and Jure Leskovec.
\newblock Graphrnn: Generating realistic graphs with deep auto-regressive models, 2018.

\end{thebibliography}

\end{document}